\documentclass[letterpaper,journal]{IEEEtran}
\usepackage{cite}
\usepackage{amsmath,amssymb,amsfonts}
\usepackage{mathtools}
\usepackage{amsthm}
\theoremstyle{definition}
\newtheorem{assumption}{Assumption}
\newtheorem{remark}{Remark}
\newtheorem{theorem}{Theorem}
\newtheorem{lemma}{Lemma}
\newtheorem{definition}{Definition}

\usepackage{algorithmic}
\usepackage{graphicx}
\usepackage{algorithm}
\usepackage{hyperref}
\usepackage{textcomp}
\usepackage{stfloats}
\usepackage{url}
\usepackage{verbatim}
\usepackage{booktabs}
\IfFileExists{SupplementaryMaterial.aux}{%
  \usepackage{xr-hyper}
  \externaldocument{SupplementaryMaterial}
}{}
\usepackage[caption=false,font=normalsize,labelfont=sf,textfont=sf]{subfig}
\begin{document}

\title{Alpha-Beta HMM: Interpretable Low-Parameter Hidden Markov Filtering in Dynamic Environments}

\author{Dongyan Sui, \IEEEmembership{Student Member, IEEE}, Haotian Pu, Yulin Shao, \IEEEmembership{Member, IEEE}, Siyang Leng, \IEEEmembership{Member, IEEE} and Stefan Vlaski, \IEEEmembership{Member, IEEE}
\thanks{This work was conducted while D.S. was a visiting PhD student at Imperial College London with support from the China Scholarship Council (No. 202406100217). S.L. was supported by the National Natural Science Foundation of China (No. 12471457). (Corresponding author: Siyang Leng)}
\thanks{Dongyan Sui is with College of Intelligent Robotics and Advanced Manufacturing, Fudan University, Shanghai, China (e-mail: dysui22@m.fudan.edu.cn). }
\thanks{Haotian Pu is with Research Institute of Intelligent Complex Systems, Fudan University, Shanghai, China (e-mail: htpu20@fudan.edu.cn). }
\thanks{Yulin Shao is with Department of Electrical and Computer Engineering, the University of Hong Kong, Hong Kong S.A.R., China (e-mail: ylshao@hku.hk).}
\thanks{Siyang Leng is with College of Intelligent Robotics and Advanced Manufacturing and Research Institute of Intelligent Complex Systems, Fudan University, Shanghai, China (e-mail: syleng@fudan.edu.cn).}
\thanks{Stefan Vlaski is with Department of Electrical and Electronic Engineering, Imperial College London, UK (e-mail: s.vlaski@imperial.ac.uk).}}

\markboth{IEEE TRANSACTIONS ON SIGNAL PROCESSING}%
{SUI \MakeLowercase{\textit{et al.}}: Alpha-Beta HMM: Hidden Markov Model Filtering with Equal Exit Probabilities and a Step-Size Parameter}


\maketitle

\begin{abstract}
Practical online inference in dynamic environments requires a lightweight filtering mechanism that remains adaptive to state changes while retaining reliable information from past noisy observations. To address this challenge, we propose the $\alpha\beta$-HMM, an interpretable low-parameter hidden Markov filtering framework that replaces the full transition matrix with an equal-exit surrogate governed by an exit-probability parameter $\alpha$, and introduces a step-size parameter $\beta$ through a generalized measurement update to regulate the influence of observational evidence. A central feature of the proposed method is that it preserves the nonlinear log-belief-ratio dynamics of HMM-type filtering, which turn out to be critical for strong performance. To analyze this nonlinear recursion, we develop a dynamical-systems framework and a deterministic reference system, through which we characterize adaptation capability, learning performance, and practical guidance for selecting the two proposed parameters. In parallel, we study the approximation error induced by the equal-exit surrogate and show that the resulting low-parameter filter remains competitive with the oracle HMM across a broad range of environments. These results reveal an explicit learning-adaptation trade-off induced by the two proposed parameters, provide principled guidance for parameter tuning, and show that strong filtering performance can be achieved within a tractable and interpretable low-parameter framework.
\end{abstract}


\section{Introduction}
\label{sec:introduction}
\IEEEPARstart{B}{ayesian} filtering is a fundamental tool for sequential inference in dynamic environments, where an agent aims to infer a latent state from noisy observations collected over time. This problem arises in a broad range of signal processing and learning applications, including speech recognition~\cite{18626,10.1561/2000000004}, target tracking~\cite{107416}, activity recognition~\cite{223161}, and online inference based on trained machine learning models~\cite{9733186,9633213}. When the underlying state remains fixed, recursive Bayesian updates provide a natural mechanism for accumulating evidence over time~\cite{978374,sarkka2023bayesian,jadbabaie2012non}. In many practical settings, however, the environment is dynamic: the true state may evolve, sometimes abruptly, and past observations can progressively lose relevance to the present. An effective online filter must therefore balance two competing goals: retaining useful memory in stable periods while remaining adaptive to state changes.

A classical framework for dynamic inference is the hidden Markov model (HMM), in which the latent state evolves according to a Markov chain and the observations are generated conditionally on the current state. Despite its importance, this formulation presents two major difficulties in many online settings. First, for an $M$-state model, a general HMM requires an $M\times M$ transition matrix, whose reliable estimation can be difficult when transition data are limited or when a large value of $M$ is needed for fine discretization. Second, the full transition matrix obscures the separate roles of environmental volatility, observation quality, and model accuracy, which makes both interpretation and principled tuning difficult. Consequently, there remains a gap between the expressive but parameter-heavy HMM filter and simpler low-complexity updates that are easier to interpret but suboptimal in changing environments.

To address this gap, we introduce the $\alpha\beta$-HMM, a low-dimensional surrogate filtering framework that captures the problem from two complementary angles using two interpretable parameters. To model environmental volatility, we replace the unknown full transition matrix by an equal-exit surrogate and introduce a scalar exit probability $\alpha$, which summarizes the tendency of the environment to leave the current state. To model observation quality and likelihood model accuracy, we introduce a generalized measurement update with a step-size parameter $\beta$, which controls the influence of incoming evidence relative to retained memory. A key feature of the proposed framework is that it preserves the nonlinear log-belief-ratio dynamics of HMM-type filtering while drastically simplifying the transition model. As will be shown later, this preserved nonlinearity plays a much more significant role in filtering performance than the equal-exit surrogate approximation itself.

Our work is related to several lines of prior research. Classical studies on HMM filtering have extensively investigated stability, forgetting, and observability properties under known transition dynamics~\cite{shue1998exponential,DOUC20091235,van2009observability,10.1214/08-AAP576,10209206,mcdonald2020exponential}. These works explain why HMM filters forget incorrect initial conditions, but they typically assume that the full transition mechanism is available, whereas our goal is to construct and analyze a low-dimensional surrogate when this transition structure is only partially known.

A second line of work concerns distributed inference in nonstationary environments. In such settings, the accumulation of past evidence can make recursive Bayesian updates too inertial, which has motivated constant-step-size belief updates. The adaptive social learning (ASL) framework~\cite{9475073} provides an analytically tractable linearized log-belief-ratio recursion that captures the trade-off between steady-state accuracy and adaptation speed. Relative to ASL and related linearized updates, the proposed $\alpha\beta$-HMM retains a nonlinear filtering structure in the log-belief domain and, when $\beta=1$, preserves an HMM interpretation under the equal-exit surrogate, thereby maintaining a closer connection to classical dynamic-state filtering. Distributed HMM models over graphs~\cite{kayaalp2022distributed} also enable collaborative online filtering across agents, but they still rely on rich transition information shared across the network. By contrast, the proposed $\alpha\beta$-HMM replaces the full transition matrix with a parsimonious surrogate governed by two interpretable parameters, which yields a lower-parameter formulation with stronger theoretical tractability and clearer parameter interpretability.

A third related direction concerns low-information sequential inference under partially known dynamics. Recent work on quickest change detection in Markov and hidden Markov models highlights the importance of sequential decision procedures that do not rely on a fully known dynamic model~\cite{10759851}. Although that line focuses on detection rather than recursive filtering, it reinforces the need for parsimonious online inference mechanisms in dynamic environments.

Against this background, this paper occupies a middle ground between two existing paradigms. Relative to standard HMM filtering, it replaces the full transition matrix by an interpretable surrogate with much lower parameter complexity. Relative to linearized adaptive updates, it preserves the nonlinear structure that is critical for strong performance. The resulting framework therefore combines competitive filtering performance, explicit interpretability, and strong analytical tractability in a single model.

The main contributions are summarized as follows:
\begin{itemize}
    \item We propose the $\alpha\beta$-HMM algorithm for online filtering in dynamic environments. The proposed framework replaces a full $M\times M$ transition model with two interpretable parameters while preserving the nonlinear log-belief-ratio dynamics of HMM-type filtering. In this way, it achieves a favorable balance between parameter efficiency, interpretability, and filtering performance, with clear potential for adaptive and multi-agent extensions.
    
    \item We study the rationality, approximation error, and practical competitiveness of the equal-exit surrogate. The analysis clarifies how the surrogate modeling error depends on heterogeneity in the exit probabilities and anisotropy in the destination distributions, while the experiments show that the resulting low-parameter nonlinear filter remains competitive with the oracle HMM across a broad range of environments. These results further indicate that the loss caused by the equal-exit surrogate is often limited relative to the larger degradation induced by linearization.
    
    \item We develop a nonlinear dynamical analysis of the proposed recursion through a deterministic reference system that approximates the original random dynamics. This analysis yields explicit results on convergence, adaptation, learning performance, and error probability, showing that a nonlinear HMM-type filtering model can retain much of the interpretability and tuning transparency typically associated with linearized adaptive recursions.
\end{itemize}


\section{Problem Setup and Development of the Algorithm}
\label{sec:problemsetup}
\subsection{Notation}

Throughout this paper, $i$ denotes the time index, boldface notation is used for random quantities, and $\mathbb{P}[\cdot]$, $\mathbb{E}[\cdot]$, and $\mathrm{Var}[\cdot]$ denote the probability, expectation, and variance operators, respectively. The symbol $\propto$ denotes proportionality up to a normalization constant. Specifically, for any function $f$ defined on $\Theta$, the notation $\mu(\theta)\propto f(\theta)$ means
$\mu(\theta)=\frac{f(\theta)}{\sum_{\theta'\in\Theta} f(\theta')}$. Vector inequalities and functions applied to vectors are understood element-wise. The $L_p$-norm of a vector is denoted by $\|\cdot\|_p$. Finally, $0_N$ and $1_N$ denote the $N$-dimensional column vectors whose entries are all $0$ and $1$, respectively.




\subsection{Problem Formulation and Assumptions}
We consider the Bayesian filtering problem, where a sensor provides noisy observations or signals $\boldsymbol{\xi}_{i}$ of the evolving state at each time step $i=1,2,\ldots$. The aim is to estimate the underlying true state $\boldsymbol{\theta}_i^{\star}$ at each time instant $i$ given the observations $\{\boldsymbol{\xi}_\tau\}_{\tau\le i}$. 

For simplicity, we assume that the true state $\boldsymbol{\theta}_i^{\star}$ belongs to a discrete set of $M$ possible states $\Theta = \{\theta_0, \theta_1, \ldots, \theta_{M-1}\}$. At each time step $i$, we assign to every state $\theta\in\Theta$ a belief $\boldsymbol{\mu}_{i}(\theta)$, which characterizes the confidence that $\theta$ is the underlying true state at time $i$ based on previous observations, i.e.,
\begin{equation}\label{belief}
\boldsymbol{\mu}_{i}(\theta)=\mathbb{P}\left[\theta_i^{\star}=\theta|\boldsymbol{\xi}_1,\ldots,\boldsymbol{\xi}_i\right],\quad \theta\in\Theta.
\end{equation}
Correct learning is said to occur at time $i$ if the belief $\boldsymbol{\mu}_i(\theta)$ is maximized at the true state $\theta = \boldsymbol{\theta}_i^{\star}$. Conditioned on \(\boldsymbol{\theta}_i^{\star} \), the observations $\boldsymbol{\xi}_i$ are independent and identically distributed (i.i.d.) over time $i$, take values in the space $\Xi$, and follow the probability density function $f(\cdot|\boldsymbol{\theta}_i^{\star})$. 

The observation model, which specifies the likelihood of an observation $\xi\in\Xi$ for each possible state $\theta \in \Theta$ is denoted by $L(\xi|\theta)$. Viewed as a function of $\xi$, $L(\xi|\theta)$ can be either a probability density function or a probability mass function, depending on whether $\xi$ is continuous or discrete.

To ensure that the underlying true state can be successfully learned, we impose the following typical assumptions:
\begin{assumption}[Finiteness of KL Divergence]
\label{asmp1}
For each pair of distinct states $\theta$ and $\theta'$, the Kullback–Leibler (KL) divergence\cite{thomas2006elements} between $L(\xi | \theta)$ and $L(\xi | \theta')$ satisfies:
\begin{equation}
D_{\mathrm{KL}}(L(\xi|\theta)||L(\xi|\theta'))<\infty.
\end{equation}
\end{assumption}

This assumption excludes degenerate cases in which two likelihood models are perfectly separated, so that one state can be ruled out with infinite confidence from a single observation.

\begin{assumption}[Identifiability of the Underlying True State]
\label{asmp2}
\begin{equation}
\left\{\theta_i^{\star}\right\}=\Theta_i^{\star}=\underset{\theta\in\Theta}{\arg\min}D_{\text{KL}}(f(\cdot|\theta_i^{\star})||L(\cdot|\theta)).
\end{equation}
\end{assumption}

Assumption \ref{asmp2} ensures that the underlying true state is the single best match for the observations under the likelihood model, which guarantees that successful learning can be achieved using likelihood model and observations.

\subsection{From HMM Filtering to the Equal-Exit Surrogate}

In a static environment, where the underlying true state remains constant over time, the belief \eqref{belief} can be computed exactly and recursively using Bayes' rule, resulting in the following update rule for each possible state indexed by $m$:
\begin{equation}\label{Bayes}
\boldsymbol{\mu}_i(\theta_m)\propto\boldsymbol{\mu}_{i-1}(\theta_m)L(\boldsymbol{\xi}_i|\theta_m).
\end{equation}
Such constructions are widely applied in decision-making, learning, and inference tasks for single or multi-agent systems in static environments. In multi-agent or social networks, the process where agents infer the true state by observing the decisions of peers and signals generated by the environment, is referred to as (non-)Bayesian social learning\cite{banerjee1992simple,bikhchandani1992theory,acemoglu2011bayesian,jadbabaie2012non,molavi2018theory,nedic2017fast,lalitha2018social,bordignon2022learning, wu2023frequentist}. These models are widely used in economics, political science, and sociology to model the behavior of financial markets, public opinion, and social networks\cite{chamley2004rational,acemoglu2011opinion}. 

In dynamic environments, where the true state evolves over time, the reliance of \eqref{Bayes} on past observations can make it less effective. In such scenarios, optimal filtering methods include the Kalman filter\cite{anderson2005optimal} and the hidden Markov filter\cite{krishnamurthy2016partially}. In this paper, we focus on filtering models related to the hidden Markov framework. Assuming that the transition probability matrix of the Markov chain between the $M$ states at each iteration is denoted as $P = \left[p_{nm}\right]_{M \times M}$, where $p_{nm} = \mathbb{P}\left[\theta_i^{\star} = \theta_m | \theta_{i-1}^{\star} = \theta_n \right]$, then the belief update, based on the hidden Markov model (HMM), is given by:
\begin{equation}\label{hmm}
\boldsymbol{\mu}_i(\theta_m)\propto\sum_{n=0}^{M-1}p_{nm}\boldsymbol{\mu}_{i-1}(\theta_n)L(\boldsymbol{\xi}_i|\theta_m).
\end{equation}

Observe that the HMM belief update depends on the entries of the full state transition matrix $P$. In practice, however, this matrix is often unavailable or difficult to estimate reliably. In many online settings, one may know that the environment is dynamic, while lacking sufficient information to identify the detailed transition probabilities between all pairs of states. Motivated by this consideration, we study a low-dimensional surrogate HMM update obtained under an equal-exit assumption for state transitions. Specifically, we assume that the true state exits its current value with a uniform probability across all alternative states. Under this assumption, the transition matrix simplifies to
\begin{equation}
 p_{mn}=\begin{cases}
1-\alpha(M-1), & m=n, \\
\alpha, & m\neq n .
\end{cases}
\end{equation}
where $\alpha$ is the probability of transitioning from the current state to any particular different state. Equivalently, the total probability of leaving the current state is $(M-1)\alpha$, so that $\alpha$ quantifies the overall volatility of the underlying true state. Substituting this transition model into the standard HMM recursion yields
\begin{equation}\label{alpha_hmm_og}
\boldsymbol{\mu}_{i}(\theta_m)\propto\left((1-\alpha M)\boldsymbol{\mu}_{i-1}(\theta_m)+\alpha\right)L(\boldsymbol{\xi}_{i}|\theta_m),
\end{equation}
For a general transition matrix $P$, the update rule \eqref{alpha_hmm_og} is generally suboptimal relative to the HMM filter \eqref{hmm}. Its advantage, however, is that it replaces the full $M\times M$ transition model with a single scalar parameter $\alpha$, thereby substantially reducing the burden of parameter specification and enabling a more transparent analysis of parameter tuning and adaptation behavior. In particular, \eqref{alpha_hmm_og} shows that $\alpha$ directly controls how much weight is retained on prior beliefs relative to the most recent observation. When $\alpha=0$, the recursion reduces to the classical Bayesian update \eqref{Bayes}. Since our focus is on filtering in dynamic environments, we restrict attention to the case $\alpha>0$ in the sequel.

By definition, in a more volatile environment, the value of $\alpha$ should be higher, indicating a greater likelihood of state transitions. Conversely, in a relatively stable environment, $\alpha$ should be smaller, indicating that the system is more likely to remain in the current state. In particular, when $\alpha = 1/M$, Eq.~\eqref{alpha_hmm_og} reduces to
$\boldsymbol{\mu}_{i}(\theta_m) \propto L(\boldsymbol{\xi}_i | \theta_m)$.
In this case, the filter completely loses memory of past observations and relies solely on the current observation for learning.

\subsection{Generalized Measurement Update and the \texorpdfstring{$\alpha\beta$-HMM}{alpha-HMM}}

The HMM update can also be interpreted as a two-step procedure: a prediction step based on the Chapman-Kolmogorov equation followed by a measurement update,
\begin{subequations}\label{two_step_update}
\begin{align}
\tilde{\boldsymbol{\mu}}_{i}(\theta_m)
&=\sum_{n=0}^{M-1}p_{nm}\boldsymbol{\mu}_{i-1}(\theta_n),
&&\text{(prediction)}
\label{prediction}\\
\boldsymbol{\mu}_{i}(\theta_m)
&\propto\tilde{\boldsymbol{\mu}}_{i}(\theta_m)L(\boldsymbol{\xi}_{i}|\theta_m),
&&\text{(measurement)}
\label{measurement}
\end{align}
\end{subequations}
Moreover, the measurement step above can be viewed as a special case of the following variational problem:
\begin{equation}
\label{optimization_generalized}
\mu_i=\arg\min_f\Bigl\{D_{\mathrm{KL}}(f\|\tilde{\mu}_i)-\beta\,\mathbb{E}_f[\log L(\boldsymbol{\xi}_i|\theta)]\Bigr\},
\end{equation}
where the standard Bayesian update in \eqref{measurement} is recovered when $\beta=1$. This formulation corresponds to a generalized Bayesian, or tempered Bayesian, inference rule, which is well established in statistical learning as a principled way to control the influence of data relative to prior information \cite{McAllester1999PACBayes,Catoni2007PACBayes,FrielPettitt2008PowerPosterior,Bissiri2016GeneralBayes}. In particular, such a temperature or learning-rate parameter is often interpreted as a robustness mechanism: when observations are noisy or the likelihood model is imperfect, choosing $\beta<1$ can prevent overly aggressive posterior concentration, while larger $\beta$ places more emphasis on new data \cite{Grunwald2012SafeBayes,MillerDunson2019Coarsening}.

Combining this generalized measurement update with the equal-exit prediction model introduced above, the resulting filtering algorithm considered in this paper can be written as
\begin{equation}
\label{alpha_hmm}
\boldsymbol{\mu}_{i}(\theta_m)\propto\left((1-\alpha M)\,\boldsymbol{\mu}_{i-1}(\theta_m)+\alpha\right)L^\beta(\boldsymbol{\xi}_{i}|\theta_m).
\end{equation}
The two parameters in the proposed framework play complementary roles: $\alpha$ modulates the prediction step and captures the volatility of the environment, whereas $\beta$ modulates the measurement step and controls the influence of observational evidence. Together, they provide a simple yet expressive mechanism for regulating the filtering process in dynamic environments. In this paper, we study the steady-state behavior of the resulting $\alpha\beta$-HMM recursion \eqref{alpha_hmm} through a nonlinear dynamical-systems framework, quantify the learning-adaptation trade-off induced by $\alpha$ and $\beta$, and provide guidance for their principled tuning.

\begin{figure*}[!t]
\centerline{\includegraphics[width=2\columnwidth]{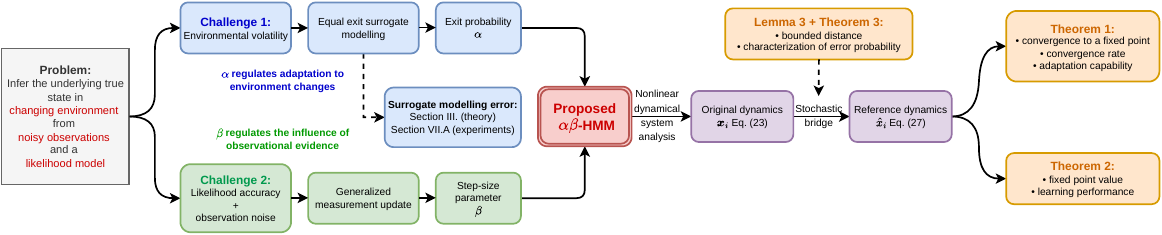}}
\caption{Overall logic of the proposed $\alpha\beta$-HMM framework.}
\label{fig:framework_logic}
\end{figure*}

A schematic summary of these relationships and the overall logic of the proposed framework is provided in Fig.~\ref{fig:framework_logic}.

\section{Approximation Error Induced by the Equal Exit Surrogate}
\label{sec:approximationerror}
The previous section motivates the proposed $\alpha\beta$-HMM as a low-dimensional surrogate of the HMM filter when the full transition matrix is unavailable or difficult to estimate. A natural question is then whether this surrogate remains accurate when the true transition matrix does not satisfy equal exit probabilities. In this section, we address this question by quantifying the approximation error induced by replacing a general transition matrix with its equal-exit counterpart. Our goal is to identify structural conditions under which the proposed surrogate is accurate and to provide a theoretical guarantee on the resulting filtering error.

\subsection{Best Equal-Exit Approximation of a General Transition Matrix}

Consider a general row-stochastic transition matrix $P^\star=[p^\star_{mn}]_{M\times M}$. We compare it with the equal-exit family $\tilde P(\alpha)=\big[\tilde p_{mn}(\alpha)\big]_{M\times M}$, where
\begin{equation}
\tilde p_{mn}(\alpha)=
\begin{cases}
1-(M-1)\alpha, & m=n,\\
\alpha, & m\neq n.
\end{cases}
\end{equation}
We first identify the equal-exit matrix that best approximates $P^\star$, namely,
\begin{equation}
\label{alpha_star_def}
\alpha^\star=\underset{\alpha}{\arg\min}\;\left\|P^\star-\tilde P(\alpha)\right\|_F^2.
\end{equation}
A direct calculation yields
\begin{equation}
\label{optimal_alpha}
\alpha^\star=\frac{\bar h}{M-1},\quad \bar h=\frac{1}{M}\sum_{m=0}^{M-1}\sum_{n\neq m}p_{mn}^\star.
\end{equation}
where $\bar h$ is the average exit probability across all states. The derivation is provided in Supplementary Material ~I. Intuitively, the choice of $\alpha^\star$ preserves the average rate at which the true state leaves its current value and therefore captures the dominant volatility level encoded by $P^\star$.

To further characterize the discrepancy between $P^\star$ and its equal-exit approximation $\tilde P(\alpha^\star)$, for each state $\theta_m$ and $n\neq m$, we define
\begin{equation}
q_m(n)\triangleq\frac{p^\star_{mn}}{h_m},\quad h_m=\sum_{n\neq m}p_{mn}^\star.
\end{equation}
where $h_m$ is the total probability of leaving state $\theta_m$, and $q_m$ is the corresponding destination distribution conditioned on exiting $\theta_m$. The mismatch between the true transition law and the equal-exit surrogate can then be decomposed into two components: heterogeneity in the exit probabilities $\{h_m\}$ and anisotropy in the destination distributions $\{q_m\}$. Specifically, we introduce
\begin{equation}
\eta_{\rm exit}\triangleq\max_m |h_m-\bar h|,
\quad
\eta_{\rm dis}\triangleq\max_m \|q_m-u_m\|_1,
\end{equation}
where $u_m$ denotes the uniform distribution over the $M-1$ states distinct from $\theta_m$. These two quantities provide interpretable measures of the discrepancy between the true transition matrix and the equal-exit surrogate, and their importance will be further illustrated through numerical experiments in Section~\ref{sec:numerical}.

\subsection{Surrogate-Induced Filtering Error Bound}

Since modifying the transition matrix affects only the prediction step in \eqref{two_step_update}, we now quantify how the mismatch between the true transition matrix and its equal-exit approximation influences the prediction stage of the filtering recursion. Let $P^\star$ denote the true transition matrix, and let $\tilde P(\alpha^\star)$ be its best equal-exit approximation obtained in the previous subsection. Define $\Delta_F \triangleq \|P^\star-\tilde P(\alpha^\star)\|_F$. Let $\mu_{i-1}^\star$ and $\mu_{i-1}^{\rm eq}$ denote the posterior distributions at time $i-1$ generated by the HMM filter with transition matrix $P^\star$ and by the equal-exit surrogate filter, respectively. Their corresponding predicted beliefs are
\begin{equation}
\hat{\mu}_i^\star=(P^\star)^\top\mu_{i-1}^\star,\quad
\hat{\mu}_i^{\rm eq}=\tilde P(\alpha^\star)^\top\mu_{i-1}^{\rm eq}.
\end{equation}

A direct decomposition yields
\begin{equation}
\hat{\mu}_i^\star-\hat{\mu}_i^{\rm eq}
=
(P^\star)^\top(\mu_{i-1}^\star-\mu_{i-1}^{\rm eq})
+
\left((P^\star)^\top-\tilde P(\alpha^\star)^\top\right)\mu_{i-1}^{\rm eq}.
\end{equation}
The first term reflects the propagation of the posterior discrepancy from the previous time step, whereas the second term captures the additional modeling error introduced by the equal-exit surrogate.

Taking the $L_1$-norm on both sides, and using the fact that $P^\star$ is row-stochastic and $\mu_{i-1}^{\rm eq}$ is a probability vector, we obtain
\begin{equation}
\|\hat{\mu}_i^\star-\hat{\mu}_i^{\rm eq}\|_1
\le
\|\mu_{i-1}^\star-\mu_{i-1}^{\rm eq}\|_1
+
\sqrt{M}\Delta_F.
\end{equation}
This bound shows that the one-step prediction error induced by the equal-exit surrogate is explicitly controlled by the transition mismatch $\Delta_F$. In particular, when the true transition matrix is well approximated by its equal-exit projection, the surrogate introduces only a limited perturbation at the prediction stage, which in turn suggests a small performance gap. This point will be further illustrated by the numerical experiments in Section~\ref{sec:numerical}.

\section{Steady State Dynamical Analysis}
\label{sec2}
While we motivated the \( \alpha \beta\)-HMM algorithm from a Markov chain with equal exit probabilities, we are interested in understanding its learning dynamics in a broader range of settings with potentially time-varying states. To this end, we will begin by quantifying in detail the ability of the \( \alpha \beta \)-HMM algorithm to identify the underlying state of the environment when it remains constant for an extended period of time. We will then proceed to characterize the adaptation capabilities of the algorithm when the environment does make a sudden change. Thus, without loss of generality, we assume that the underlying true state is $\theta_0 \in \Theta$, i.e. $\theta_i^{\star}=\theta_0$ for all $i=1,2,\ldots$. We aim to analyze the dynamics of the log-likelihood ratio of the belief on wrong and true states, i.e.:
\begin{equation}
\boldsymbol{x}_{m,i}\triangleq \log\frac{\boldsymbol{\mu}_{i}(\theta_m)}{\boldsymbol{\mu}_{i}(\theta_0)},\quad m=1,\ldots,M-1,\quad i=0,1,\ldots.
\end{equation}

We refer to $\boldsymbol{x}_{m,i}$ as the log-belief ratio. Using \eqref{alpha_hmm}, the log-belief ratio evolves as:
\begin{equation}
\boldsymbol{x}_{m,i}=\log\frac{(1-\alpha M)\boldsymbol{\mu}_{i-1}(\theta_m)+\alpha}{(1-\alpha M)\boldsymbol{\mu}_{i-1}(\theta_0)+\alpha}+\beta\log\frac{L(\boldsymbol{\xi}_i|\theta_m)}{L(\boldsymbol{\xi}_i|\theta_0)}.
\end{equation}

The above dynamical system is both \emph{nonlinear} and \emph{stochastic}---\emph{nonlinear} since the first term of the r.h.s. is nonlinear with respect to $\boldsymbol{x}_{m,i-1}$ and \emph{stochastic} since the system is driven by the function of a random process $\boldsymbol{\xi}_i$.

We first characterize the nonlinear mapping of the log-belief ratio from one time-instance to the next. From Eq.~\eqref{alpha_hmm}, we have $\sum_{m=0}^{M-1}\boldsymbol{\mu}_{i}(\theta_m) = 1$, for all $i = 0, 1, \ldots$. Utilizing this property, we can expand:
\begin{equation}\label{nonlinearfunction}
\begin{aligned}
&\log\frac{(1-\alpha M)\boldsymbol{\mu}_{i-1}(\theta_m)+\alpha}{(1-\alpha M)\boldsymbol{\mu}_{i-1}(\theta_0)+\alpha}\\
=&\log\frac{(1-\alpha M)\boldsymbol{\mu}_{i-1}(\theta_m)+\alpha\sum_{n=0}^{M-1}\boldsymbol{\mu}_{i-1}(\theta_n)}{(1-\alpha M)\boldsymbol{\mu}_{i-1}(\theta_0)+\alpha\sum_{n=0}^{M-1}\boldsymbol{\mu}_{i-1}(\theta_n)}\\
=&\log\frac{(1-\alpha M)\frac{\boldsymbol{\mu}_{i-1}(\theta_m)}{\boldsymbol{\mu}_{i-1}(\theta_0)}+\alpha+\alpha\sum_{n=1}^{M-1}\frac{\boldsymbol{\mu}_{i-1}(\theta_n)}{\boldsymbol{\mu}_{i-1}(\theta_0)}}{1-\alpha M+\alpha+\alpha\sum_{n=1}^{M-1}\frac{\boldsymbol{\mu}_{i-1}(\theta_n)}{\boldsymbol{\mu}_{i-1}(\theta_0)}}\\
=&\log\frac{(1-\alpha M)\exp(\boldsymbol{x}_{m,i-1})+\alpha+\alpha\sum_{n=1}^{M-1}\exp(\boldsymbol{x}_{n,i-1})}{1-\alpha M+\alpha+\alpha\sum_{n=1}^{M-1}\exp(\boldsymbol{x}_{n,i-1})}.
\end{aligned}
\end{equation}

Defining:
\begin{equation}\label{Fm}
\begin{aligned}
&F_{m}(x_1,\ldots,x_{M-1})\triangleq\\
&\log\frac{(1-\alpha M)\exp(x_{m})+\alpha+\alpha\sum_{n=1}^{M-1}\exp(x_{n})}{1-\alpha M+\alpha+\alpha\sum_{n=1}^{M-1}\exp(x_{n})},
\end{aligned}
\end{equation}
the dynamics for all $m=1,\ldots,M-1$ and $i=1,2,\ldots$ can be expressed compactly as:
\begin{equation}\label{stochastic}
\boldsymbol{x}_{m,i}=F_m(\boldsymbol{x}_{1,i-1},\ldots,\boldsymbol{x}_{M-1,i-1})+\beta\log\frac{L(\boldsymbol{\xi}_i|\theta_m)}{L(\boldsymbol{\xi}_i|\theta_0)}.
\end{equation}

Now we will move on to the stochastic part. The second term on the r.h.s. of \eqref{stochastic} satisfies that for all $m=1,\ldots,M-1$,
\begin{align}
&\mathbb{E}\left[\log\frac{L(\boldsymbol{\xi}_i|\theta_m)}{L(\boldsymbol{\xi}_i|\theta_0)}\right]=\int_{\xi\in\Xi}f(\xi|\theta_0)\log\frac{L(\xi|\theta_m)}{L(\xi|\theta_0)}\mathrm{d}\xi\nonumber\\
=&\int_{\xi\in\Xi}f(\xi|\theta_0)\log\frac{f(\xi|\theta_0)}{L(\xi|\theta_0)}\mathrm{d}\xi-\int_{\xi\in\Xi}f(\xi|\theta_0)\log\frac{f(\xi|\theta_0)}{L(\xi|\theta_m)}\mathrm{d}\xi\nonumber\\
=&D_{\text{KL}}(f(\cdot|\theta_0)||L(\cdot|\theta_0))-D_{\text{KL}}(f(\cdot|\theta_0)||L(\cdot|\theta_m))\nonumber\\
\triangleq & -d_m<0.
\end{align}
The last inequality above is guaranteed by Assumption \ref{asmp2}. From the definition of $d_m$, it is evident that it characterizes the filter's ability to distinguish any incorrect state $\theta_m$ from the underlying true state $\theta_0$ using the likelihood models $L(\cdot|\theta_m)$ and the observation $\boldsymbol{\xi}_i$. A larger $d_m$ indicates stronger identifiability. Define the identifiability vector $d=\left[d_1,\ldots,d_{M-1}\right]^\top$, the state vector
$\boldsymbol{x}_{i}=\left[\boldsymbol{x}_{1,i},\ldots,\boldsymbol{x}_{M-1,i}\right]^\top$, and the mapping
\begin{equation}\label{iterative_mapping}
F(x)=\left[F_1(x_1,\ldots,x_{M-1}),\ldots,F_{M-1}(x_1,\ldots,x_{M-1})\right]^\top.
\end{equation}
Taking conditional expectations and rewriting in vector form, the dynamics in \eqref{stochastic} become
\begin{equation}
\mathbb{E}\!\left[\boldsymbol{x}_{i}| \boldsymbol{x}_{i-1}\right]
=F(\boldsymbol{x}_{i-1})-\beta d.
\end{equation}
This shows that the stochastic recursion admits the deterministic drift $F(\boldsymbol{x}_{i-1})-\beta d$. Accordingly, the system \eqref{stochastic} can be written as the sum of this drift term and a zero-mean innovation term. Motivated by this decomposition, we introduce the following deterministic dynamical system, referred to as the \textbf{reference dynamical system}:
\begin{equation}\label{deterministic}
\hat{x}_{i}=F(\hat{x}_{i-1})-\beta d.
\end{equation}
Here $\hat{x}_i$ should not be interpreted as an exact recursion for $\mathbb{E}[\boldsymbol{x}_i]$; rather, it serves as the drift-matched deterministic counterpart of the stochastic log-belief-ratio dynamics.

In the following theoretical analysis, we focus on the scenario where $0 < \alpha < 1/M$, ensuring $0 < 1 - \alpha M < 1$. This setting, referred to as slow-changing conditions, corresponds to the regime where past observations are likely to be informative of the current state and also facilitates a more analytically tractable analysis of the algorithm.

\begin{remark}
\label{remark1}
It is worth noting that for all $m=1,\ldots,M-1$,
\begin{equation}
F_m(0,\ldots,0)=\:0,\quad
\frac{\partial F_m}{\partial x_n}\Big|_{(0,\ldots,0)}=\begin{cases}
1-\alpha M, & \text{if}\;n=m \\
0, & \text{if}\;n\neq m.
\end{cases}
\end{equation}
By applying a multivariate Taylor expansion to $F(x)$ around $x = 0_{M-1}$ up to the first-order term, for all $m=1,\ldots,M-1$ we have:
\begin{equation}
F_m(x_1,\ldots,x_{M-1})=(1-\alpha M)x_{m}+o(\left\|x\right\|),
\end{equation}
then we derive the linear approximation of the system \eqref{stochastic}, which is:
\begin{equation}\label{linearized_ahmm}
\boldsymbol{x}_{m,i}=(1-\alpha M)\boldsymbol{x}_{m,i-1}+\beta\log\frac{L(\boldsymbol{\xi}_i|\theta_m)}{L(\boldsymbol{\xi}_i|\theta_0)}.
\end{equation}
The above equation can be viewed as a model resembling the single-agent version of the adaptive social learning (ASL) recursion~\cite{9475073}, which takes the form
\begin{equation}\label{asl_single}
\boldsymbol{x}_{m,i}=(1-\delta)\boldsymbol{x}_{m,i-1}+\delta\log\frac{L(\boldsymbol{\xi}_i|\theta_m)}{L(\boldsymbol{\xi}_i|\theta_0)},
\end{equation}
where $\delta$ is the step-size parameter. Equations~\eqref{asl_single} and~\eqref{linearized_ahmm} both apply a discount to past information during the recursion. The key difference is that the former uses a single parameter, whereas the latter retains two independent parameters: $\alpha$, which characterizes the rate of change of the environment, and $\beta$, which controls the trade-off between retained memory and new observational evidence. As remarked earlier, this added flexibility is useful because the choice $\beta=1$ recovers the HMM filter associated with a Markov chain with equal exit probability, while setting $\alpha=\frac{\delta}{M}$ and $\beta=\delta$ yields the adaptive social learning recursion in~\cite{9475073}. Therefore, the linearized model provides a direct bridge between the proposed nonlinear $\alpha\beta$-HMM and existing adaptive belief updates. This connection is important because it isolates the role of nonlinearity, which will be shown in the numerical results to have a much stronger impact on filtering performance than the equal-exit surrogate approximation.
\end{remark}




\subsection{Convergence Rate and adaptation Ability}
In this part, we prove that the reference dynamical system \eqref{deterministic} converges to a fixed point at an exponential rate. Based on this, we can analyze the adaptation capability and how it can be influenced by the exit probability $\alpha$, step-size parameter $\beta$ and identifiability vector $d$.

\begin{definition}[Contractions]
\label{def1}
Let $(X,\rho)$ be a metric space. A mapping $T: X\to X$ is a \textbf{contraction}, if there exists a constant $c$, with $0\le c< 1$, such that
\begin{equation}
\rho\left(T(x),T(y)\right)\le c \rho(x,y)
\end{equation}
for all $x,y\in X$.
\end{definition}

Based on the above definition, we now present the following lemma, which demonstrates that the mapping $F(x)$ in~\eqref{iterative_mapping} is a contraction on the non-positive region; this property facilitates the convergence analysis of the reference dynamical system.
\begin{lemma}
\label{lemma1}
Consider $0<\alpha<1/M$ and $\beta>0$, the mapping $F$ defined in \eqref{iterative_mapping} and \eqref{Fm} has the following properties:
\begin{enumerate}
    \item $F(\left(-\infty,0\right]^{M-1})\subset\left(-\infty,0\right]^{M-1}$; that is, $F$ maps the non-positive region into itself.
    \item $F$ is a contraction on $\left(-\infty,0\right]^{M-1}$ with respect to the $L_\infty-$norm.
\end{enumerate}
\end{lemma}
\begin{proof}
See Supplementary Material~II.
\end{proof}

Let $T: X\to X$ be a mapping. A point of $x\in X$ such that $T(x)=x$ is called a \textbf{fixed point} of $X$. For a discrete dynamical system $x_i = T(x_{i-1})$, a fixed point of the map $T$ corresponds to an equilibrium state of the system. In the remainder of this paper, we will consistently refer to it as the fixed point.

For the reference dynamical system \eqref{deterministic}, its fixed point has the following property:

\begin{lemma}
\label{lemma2}
The fixed point $\hat{x}^\infty$ of the reference dynamical system \eqref{deterministic} satisfies $\hat{x}^\infty=\left[\hat{x}_1^\infty,\ldots,\hat{x}_{M-1}^\infty\right]^\top$,
\begin{equation}\label{equilibrium}
\hat{x}^\infty_m =\log\frac{\alpha}{\alpha \exp(\beta d_m)+(1-\alpha M)(\exp(\beta d_m)-1)\hat{\mu}_0^\infty},
\end{equation}
where $\hat{\mu}_0^\infty=1/\left(1+\sum_{m=1}^{M-1}\exp(\hat{x}_m^\infty)\right)$.
\end{lemma}
\begin{proof}
See Supplementary Material~III.
\end{proof}

Lemma \ref{lemma2} shows that the reference dynamical system has a fixed point, which is strictly negative. To verify this, for all $m=1,\ldots,M-1$, when $1-\alpha M > 0$ we have:
\begin{equation}
\alpha \exp(\beta d_m)+(1-\alpha M)(\exp(\beta d_m)-1)\hat{\mu}^\infty_0>\alpha\exp(\beta d_m),
\end{equation}
implying $\hat{x}_m^\infty < -\beta d_m < 0$.

This result implies that the reference dynamical system learns correctly, since the log-belief ratio $\hat{x}_{m,i}$ is strictly less than zero for any wrong state $\theta_m$ . When $1-\alpha M > 0$, increasing either $\beta$ (the weight assigned to the observational data) or $d_m$ (the distinguishability between $\theta_0$ and $\theta_m$) drives the fixed-point value $\hat{x}_m^\infty$ further negative, indicating stronger confidence in the true state relative to the competing state $\theta_m$.

Now we are able to provide the main result regarding the convergence of the reference dynamical system \eqref{deterministic} and its convergence rate.

\begin{theorem}
\label{thm1}
When $0<\alpha<1/M$ and $\beta>0$, $\hat{x}_{i}$ defined in \eqref{deterministic} will converge to a unique fixed point, i.e.,
$\lim_{i\to\infty}\hat{x}_{i}=\hat{x}^\infty$, and the convergence rate satisfies
\begin{equation}\label{thm1_result2}
\left\|\hat{x}_i-\hat{x}^\infty\right\|_\infty\le \lambda \left\|\hat{x}_{i-1}-\hat{x}^\infty\right\|_\infty,
\end{equation}
where
\begin{equation}\label{contraction_rate}
\lambda=1-\min\left\{\frac{2\alpha}{1-\alpha M+2\alpha},\frac{\beta\underline{d}}{\overline{x}_{0}-\log\frac{\alpha}{1-\alpha M+\alpha}+\beta\underline{d}}\right\},
\end{equation}
\begin{equation}
\label{dmin_def1}
\overline{x}_0=\max\left\{\left\{\hat{x}_{m,0}\right\}_{m=1}^{M-1},0\right\},\quad \underline{d}=\underset{m=1,\ldots,M-1}{\min}d_m.
\end{equation}
\end{theorem}
\begin{proof}
See Supplementary Material~IV.
\end{proof}

Theorem \ref{thm1} provides an upper bound on the convergence rate of $\hat{x}_i$ to $\hat{x}^\infty$ in the sense of the $L_\infty$-norm, which decreases as the exit probability $\alpha$ increases, as shown in \eqref{contraction_rate}. We can also observe the impact of the data, captured in $\beta \underline{d}$, and the initial setting, captured in $\overline{x}_0$, on the convergence rate of the reference dynamical system to its fixed point. The parameter $\underline{d}$ depends on both the local likelihood function (e.g., pre-trained models) and the quality of observations (e.g., variance of local data noise). This parameter, referred to as the \emph{identifiability parameter}, reflects the ability to distinguish between the two most challenging states. From Theorem \ref{thm1}, it follows that higher identifiability $\underline{d}$ and higher $\beta$ accelerate the convergence to the fixed point, while the deviation in the initial setting, captured in $\overline{x}_0$, slows down the convergence rate.

\begin{figure}[!t]
\centerline{\includegraphics[width=\columnwidth]{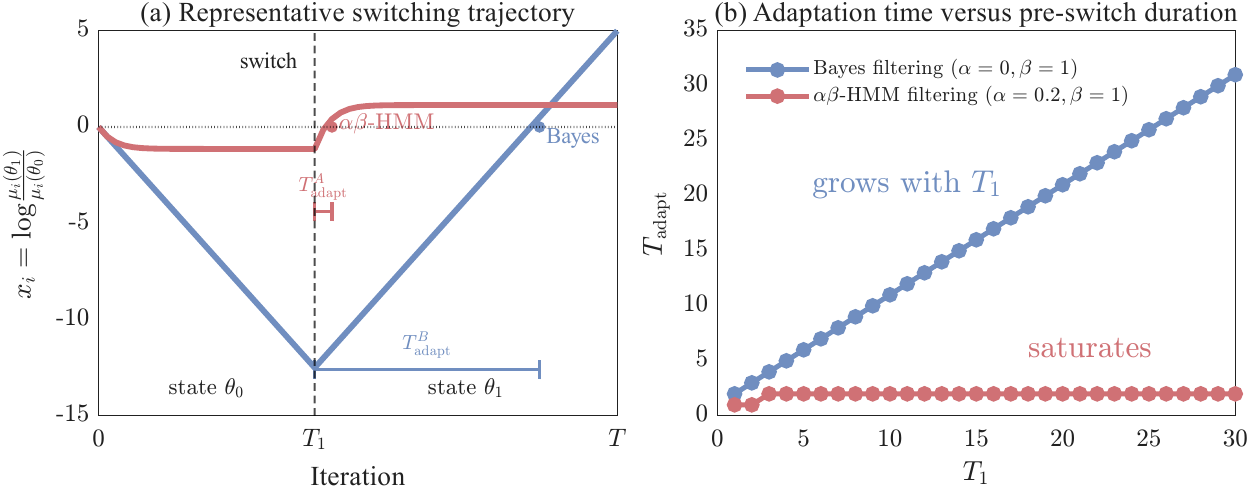}}
\caption{Illustration of the adaptation behavior of Bayes filtering and $\alpha\beta$-HMM ($\beta=1$) under a binary switching environment. Two candidate states, $\theta_0$ and $\theta_1$, are considered, and both the likelihood models and the corresponding environmental observation models are taken as Gaussian densities with distinct means. The environment remains in state $\theta_0$ until time $T_1$, after which it switches to state $\theta_1$. (a) Representative evolution of the expected log-belief ratio $x_i=\log\!\big(\mu_i(\theta_1)/\mu_i(\theta_0)\big)$ under the two filtering strategies. The horizontal brackets indicate the post-switch adaptation intervals, where $T_{\mathrm{adapt}}^A$ and $T_{\mathrm{adapt}}^{B}$ denote the numbers of iterations required by $\alpha\beta$-HMM and Bayes filtering, respectively, for the expected log-belief ratio to become positive after the change. Owing to its forgetting effect, $\alpha\beta$-HMM adapts rapidly after the switch, whereas Bayes filtering adjusts more slowly because it continues to accumulate confidence in the pre-switch state. (b) Adaptation time $T_{\mathrm{adapt}}$ as a function of the pre-switch stationary interval $T_1$. The adaptation time of Bayes filtering increases with $T_1$, while that of $\alpha\beta$-HMM saturates.}
\label{fig1}
\end{figure}

\begin{remark}
\label{remark 3}
The convergence rate of the reference dynamical system to its fixed point, as presented in Theorem \ref{thm1}, plays a crucial role in understanding the \emph{adaptation capability} of the algorithm. Specifically, when the environment remains in a fixed state for an extended period, the reference dynamical system converges to its fixed point at an exponential rate. When the environment state changes, the corresponding fixed point shifts accordingly, and the system will converge to the new fixed point exponentially fast, as illustrated in Fig.~\ref{fig1}. The convergence speed depends on the choice of the parameters $\alpha$ and $\beta$, and the quality of the data and model. When using the traditional Bayesian update \eqref{Bayes}, as indicated by equations \eqref{nonlinearfunction} and \eqref{deterministic} with $\alpha = 0$ and $\beta=1$, the corresponding deterministic dynamical system simplifies to the linear equation $\hat{x}_{i}=\hat{x}_{i-1}-d$, which shows that the log-belief ratio evolves linearly over time. This slower rate explains why the proposed algorithm exhibits superior adaptability in dynamic environments.

To further illustrate the adaptation implications of Theorem~\ref{thm1}, Fig.~\ref{fig1} compares the proposed $\alpha\beta$-HMM with the classical Bayesian update in a binary switching scenario. The figure shows that, whereas the adaptation time of the Bayesian update grows with the length of the pre-switch stationary interval, the adaptation time of $\alpha\beta$-HMM eventually saturates. This behavior is consistent with the contraction-based analysis above and highlights the improved adaptability induced by the nonlinear recursion. A detailed derivation for this example is provided in Supplementary Material~V.
\end{remark}


\subsection{Fixed Point Value and Learning Performance}
We now quantify the learning performance of the $\alpha\beta$-HMM and examine its dependence on the exit probability $\alpha$, the step-size parameter $\beta$, and the identifiability parameter $\underline{d}$. Since we have already proven that the reference dynamical system converges to the unique fixed point as described in Lemma \ref{lemma2}, the position of the fixed point will be sufficient to characterize the resulting learning performance. Due to the nonlinearity of \eqref{equilibrium}, explicitly solving for the exact fixed point value is challenging. Nevertheless, we can provide an estimate of its value. In Theorem \ref{thm2}, we derive upper and lower bounds for the counterpart of the belief $\boldsymbol{\mu}_i(\cdot)$ in the reference dynamical system.

Given that $\boldsymbol{x}_{m,i} = \log \left(\boldsymbol{\mu}_{i}(\theta_m)/\boldsymbol{\mu}_{i}(\theta_0)\right)$, 
we can obtain
\begin{equation}
\boldsymbol{\mu}_i(\theta_m)=\begin{cases}
1/\left(1+\sum_{m=1}^{M-1}\exp(\boldsymbol{x}_{m,i})\right), &  m= 0,\\
\exp(\boldsymbol{x}_{m,i})/\left(1+\sum_{m=1}^{M-1}\exp(\boldsymbol{x}_{m,i})\right), &  m\neq 0 .
\end{cases} 
\end{equation}
Since $\hat{x}_{m,i}$ can be seen as the counterpart of $\boldsymbol{x}_{m,i}$ in reference dynamical system, similarly we can define 
\begin{equation}
\hat{\mu}_i(\theta_m)=\begin{cases}
1/\left(1+\sum_{m=1}^{M-1}\exp(\hat{x}_{m,i})\right), &  m= 0,\\
\exp(\hat{x}_{m,i})/\left(1+\sum_{m=1}^{M-1}\exp(\hat{x}_{m,i})\right), &  m\neq 0 ,
\end{cases} 
\end{equation}
and then $\hat{x}_{m,i} = \log \left(\hat{\mu}_{i}(\theta_m)/\hat{\mu}_{i}(\theta_0)\right)$, for all $m = 1, \ldots, M-1$ and $i=0,1,\ldots$. Here $\hat{\mu}_i(\theta)$ can be seen as the counterpart of belief on state $ \theta $ in the reference dynamical system. By Theorem 1 and basic limit properties,
\begin{equation}
\lim_{i\to\infty}\hat{\mu}_i(\theta_m)=\hat{\mu}_m^\infty,\quad \forall m=0,\ldots,M-1,
\end{equation}
where
\begin{equation}\label{mu_inf_def}
\hat{\mu}_m^\infty=\begin{cases}
1/\left(1+\sum_{m=1}^{M-1}\exp(\hat{x}_m^\infty)\right), &  m= 0,\\
\exp(\hat{x}_m^\infty)/\left(1+\sum_{m=1}^{M-1}\exp(\hat{x}_m^\infty)\right), &  m\neq 0 .
\end{cases} 
\end{equation}

The value of $ \hat{\mu}_0^\infty $, which represents the steady-state belief on the true state in reference dynamical system, reflects the learning performance of the algorithm: the closer $ \hat{\mu}_0^\infty $ is to 1, the better the distinction between correct and incorrect states at steady state; conversely, the closer $ \hat{\mu}_0^\infty $ is to $ 1/M $, the more ambiguous the learning outcome. We provide upper and lower bound estimates for $ \hat{\mu}_0^\infty $ in the following theorem.

\begin{theorem}
\label{thm2}
When $0<\alpha<1/M$, $\beta>0$, and the underlying true state remains unchanged, the $\hat{\mu}_0^\infty$ defined in \eqref{mu_inf_def} satisfies the following bounds:
\begin{equation}\label{lb_mu0inf}
\hat{\mu}_0^\infty\ge\frac{(1-\alpha M)\exp(\beta\underline{d})-1+\alpha}{(1-\alpha M)(\exp(\beta\underline{d})-1)}\triangleq \underline{\mu},
\end{equation}
\begin{equation}
\hat{\mu}_0^\infty\le \frac{(1-\alpha M-\alpha+\alpha/\underline{\mu})\exp(\beta\overline{d})-1+\alpha}{(1-\alpha M)(\exp(\beta\overline{d})-1)}\triangleq \overline{\mu},
\end{equation}
where
$\underline{d}=\underset{m=1,\ldots,M-1}{\min}d_m$, and $\overline{d}=\underset{m=1,\ldots,M-1}{\max}d_m$.
\end{theorem}

\begin{proof}
See Supplementary Material~VI.
\end{proof}

\begin{figure}[!t]
\centerline{\includegraphics[width=\columnwidth]{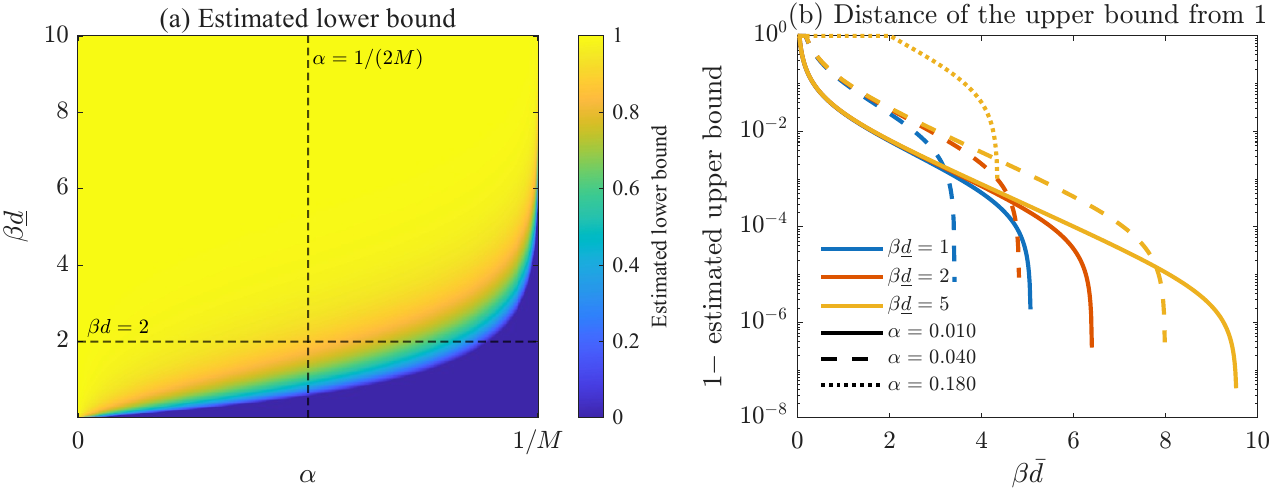}}
\caption{Visualization of the lower and upper bounds in Theorem~\ref{thm2}. (a) Lower bound as a function of $\alpha$ and $\beta\underline{d}$. The heatmap shows the lower bound over the full parameter plane, and the highlighted horizontal and vertical slices illustrate its variation when one parameter is fixed. In particular, the lower bound increases with $\beta\underline{d}$ and decreases with $\alpha$. (b) Curves of $1-\text{Upper Bound}$ versus $\beta\overline{d}$ for different values of $\beta\underline{d}$ and $\alpha$, where colors indicate $\beta\underline{d}$ and line styles indicate $\alpha$. The plot shows that the upper bound approaches one rapidly as $\beta\overline{d}$ increases, with the rate of convergence depending on both $\alpha$ and $\beta\underline{d}$.}
\label{fig2}
\end{figure}

The upper and lower bounds in Theorem \ref{thm2}, as functions of the parameters $\alpha$, $\beta\underline{d}$, and $\beta\overline{d}$, are illustrated in Figure \ref{fig2}. We observe that when $\alpha$ is smaller and $\beta\underline{d}$ is larger, the steady-state belief on the correct state in the reference dynamical system is higher, indicating improved learning performance. This observation aligns with intuition: better learning performance is expected in less volatile environments, with lower observation noise, and a more accurate local likelihood model. The step-size parameter $\beta$ and the identifiability parameter $\underline{d}$ play a decisive role; when either $\beta$ or $\underline{d}$ is large, good learning performance is ensured in deterministic dynamical system, noise-free by construction, even when $\alpha$ is relatively large. This means that, in the deterministic setting, accurate inference is possible in volatile environments, provided that the quality of the data model and observations is high enough or that the value of $\beta$ is large enough. However, choosing a large \( \beta\) comes at the cost of amplifying the effect of observation noise, as we will further demonstrate in Section~\ref{sec:numerical}.

\section{From Deterministic to Stochastic Systems}
\label{sec:3}
In this section, we discuss the relationship between the stochastic dynamical system induced by $\alpha\beta$-HMM, represented by \eqref{stochastic}, and the reference dynamical system \eqref{deterministic}, establishing that the dynamics of \eqref{deterministic} are representative of the stochastic system \eqref{stochastic}, which describes the actual learning dynamics of the \( \alpha \beta \)-HMM recursion. In Lemma \ref{lemma5}, we show that the expected distance between $\boldsymbol{x}_i$ and the fixed point of the reference dynamical system can be asymptotically bounded. Then, in Theorem \ref{thm3}, we compute the instantaneous error probability of the $\alpha\beta$-HMM based on the fixed point of the reference dynamical system. To ensure tractability of the theoretical analysis, we introduce the following assumption.

\begin{assumption}[Bounded Observation-Driven Fluctuation]
There exists a positive constant $C$ such that:
\label{asmp3}
\begin{equation}\label{C_def}
\underset{m=1,\ldots,M-1}{\max}\:\underset{\boldsymbol{\xi}\in\Xi}{\sup}\: \left|\log\frac{L(\boldsymbol{\xi}|\theta_m)}{L(\boldsymbol{\xi}|\theta_0)}+d_m\right|\le C,\quad a.s..
\end{equation}
\end{assumption}
Assumption~\ref{asmp3} is a technical regularity condition used to control the stochastic fluctuation around the deterministic reference dynamics. It is automatically satisfied for discrete or bounded-support observation models, as well as for continuous models after clipping, truncation, or quantization. We emphasize that this assumption is introduced only for deriving the explicit bound in Lemma~3 and the computable error-probability estimate in Theorem~3. It is not required for defining the proposed algorithm, nor for the deterministic steady-state analysis in Theorems~1 and Theorem~2. Common unbounded models, such as Gaussian observations on $\mathbb{R}$, generally do not satisfy this assumption globally; in such cases, the corresponding bound should be regarded as a tractable analytical surrogate rather than a universal characterization.






The following lemma establishes the relationship between the original stochastic dynamical system \eqref{stochastic} and the reference dynamical system\eqref{deterministic}.

\begin{lemma}
\label{lemma5}
Let $\boldsymbol{x}_{m,i}$ and $\hat{x}_{m,i}$ follow the dynamical system in \eqref{stochastic} and \eqref{deterministic}, respectively. Then, the following equation holds:
\begin{equation}
\mathbb{E}\left[\left\|\boldsymbol{x}_{i}-\hat{x}^\infty\right\|_\infty\right]\le O(\lambda_1^i)+\frac{\beta C}{1-\lambda_1},
\end{equation}
where $\hat{x}^\infty$ is the fixed point of reference dynamical system \eqref{deterministic},
\begin{equation}\label{lambdap}
\lambda_1=1-\min\left\{\frac{2\alpha}{1-\alpha M+2\alpha},\frac{\beta\underline{d}}{2\log\frac{1-\alpha M+\alpha}{\alpha}+\beta C}\right\},
\end{equation}
with $C$ and $\underline{d}$ defined in \eqref{C_def} and \eqref{dmin_def1}, respectively.
\end{lemma}

\begin{proof}
See Supplementary Material~VII.
\end{proof}

Lemma \ref{lemma5} shows that the expected distance between the original dynamical system and the fixed point of the reference dynamical system can be bounded. The coefficient $\lambda_1\le 1$, then as the time instant approaches infinity, the upper bound of the expected distance between the two is related to the first absolute moment of the log-likelihood ratio of the observational model and the contraction coefficient $\lambda_1$. In other words, when the observation noise decreases, the stochastic dynamical system corresponding to the $\alpha\beta$-HMM \eqref{stochastic} approaches the fixed point of the reference dynamical system.

To evaluate the learning performance of $\alpha\beta$-HMM, we introduce the \textbf{instantaneous error probability}:
\begin{equation}\label{error_prob}
p^e_{i}\triangleq \mathbb{P}\left[\;\exists\;\theta_m \neq \theta_0, \text{ s.t. } \boldsymbol{\mu}_i(\theta_m) \geq \boldsymbol{\mu}_i(\theta_0)\right].
\end{equation}
The instantaneous error probability quantifies the probability of failing to correctly identify the underlying true state at a given time $i$ during the online filtering process. We will then provide an estimate of this metric in the steady-state scenario for the proposed $\alpha\beta$-HMM algorithm.

\begin{theorem}
\label{thm3}
When $0<\alpha < 1/M$, $\beta>0$, and the underlying true state remains constant, i.e. $\theta_i^{\star}= \theta_0$ for all $i=1,2,\ldots$, the instantaneous error probability for the $\alpha\beta$-HMM algorithm \eqref{alpha_hmm} satisfies:
\begin{equation}\label{ep_ineq}
p_i^e\le\frac{O(\lambda_1^i)+\beta C}{-(1-\lambda_1)\overline{x}^\infty},
\end{equation}
where $\overline{x}^\infty=\underset{m=1,\ldots,M-1}{\max}\hat{x}_m^\infty<0$, other variable definitions remain unchanged from those presented in Lemma \ref{lemma5}. Furthermore, we have the following asymptotic estimate of the instantaneous error probability:

If $\beta=O(\alpha)$, then:
\begin{equation}\label{asymptotic}
\lim_{\alpha\rightarrow 0}\left(\underset{i\to\infty}{\lim\sup}\;p_i^e\right)=0.
\end{equation}
\end{theorem}

\begin{proof}
See Supplementary Material~VIII.
\end{proof}

The contraction coefficient $\lambda_1$ in Theorem \ref{thm3} is always less than 1, consequently the upper bound of the instantaneous error probability decays exponentially with time and eventually converges to a fixed value $\frac{\beta C}{-(1-\lambda_1)\overline{x}^\infty}$.

This result indicates that under the $\alpha\beta$-HMM framework, the probability of erroneous learning cannot be guaranteed to asymptotically approach zero for all choices of $\alpha$ and $\beta$, even in steady-state conditions, which contrasts with the behavior predicted by Bayes' formula \cite{lalitha2018social}. However, by sacrificing some learning performance, the $\alpha\beta$-HMM framework significantly enhances adaptability, as previously discussed. The steady-state error learning probability can be reduced either by (1) pushing the fixed point $\hat{x}^\infty$ of the deterministic system further away from zero and accelerating the convergence rate (i.e., reducing $\lambda_1$), noting that, as shown in earlier analyses, improving the identifiability parameter $\underline{d}$ facilitates this outcome, or by (2) reducing the observation noise, thereby increasing the informativeness of the observations.

The asymptotic analysis results for small $\alpha$ further indicate that as $\alpha$ approaches zero, the instantaneous error probability diminishes over time. This phenomenon highlights the trade-off between the learning capability and the adaptation ability of the algorithm: while a small $\alpha$ yields good learning performance, it reduces the adaptation rate, as illustrated in \eqref{lambdap}. Moreover, when $\alpha = 0$, the proposed algorithm degenerates into Bayes' formula, which, as discussed in Remark \ref{remark 3}, exhibits poor adaptability.





\section{Parameter Selection and Tuning}

The theoretical results provide a simple practical guideline for choosing the two parameters. Since they act on different parts of the recursion, their roles are complementary: $\alpha$ controls the amount of memory retained from past beliefs through the prediction step, whereas $\beta$ controls the strength of observational evidence through the measurement step. Accordingly, a practical tuning strategy should first determine $\alpha$ based on the anticipated rate of environmental change, and then determine $\beta$ based on the reliability of the observations and the accuracy of the likelihood model.

The parameter $\alpha$ should be selected according to the volatility level of the environment. Larger $\alpha$ improves responsiveness to state changes but reduces steady-state confidence, whereas smaller $\alpha$ is preferable in slowly varying or nearly static environments, where stronger memory and higher steady-state accuracy are more important. In this sense, $\alpha$ governs the balance between memory retention and adaptation capability.

The parameter $\beta$ should be selected according to the quality of the observational information. Larger $\beta$ places more weight on new observations and is preferable when the data are clean and informative and the likelihood model is accurate. Smaller $\beta$ is more conservative and more robust under noise or likelihood-model mismatch, since it tempers the influence of unreliable evidence. In practice, $\beta=1$ is a natural baseline, as it recovers the HMM-type measurement update associated with the equal-exit surrogate without additional tempering.

The two parameters should also be interpreted jointly. A larger $\alpha$ and a larger $\beta$ both increase responsiveness, but through different mechanisms: the former acts by faster forgetting of past beliefs, whereas the latter acts by stronger amplification of new evidence. Hence, in rapidly varying but relatively clean environments, choosing both a larger $\alpha$ and a larger $\beta$ can be beneficial. In noisier settings, however, it is often preferable to combine a larger $\alpha$ with a moderate or smaller $\beta$.

In practice, one may first choose $\alpha$ to match the anticipated volatility level of the environment, and then choose $\beta$ according to observation quality and model reliability. Since $\alpha$ admits a direct interpretation as an exit-probability surrogate, it is also amenable to online estimation or adaptive updating; a detailed treatment of this direction is beyond the scope of the present paper and is studied in our companion work~\cite{11464476}.

\section{Numerical Experiments}
\label{sec:numerical}
\subsection{Surrogate Error Under Different Transition Structures}

\begin{figure*}[!t]
\centerline{\includegraphics[width=1.8\columnwidth]{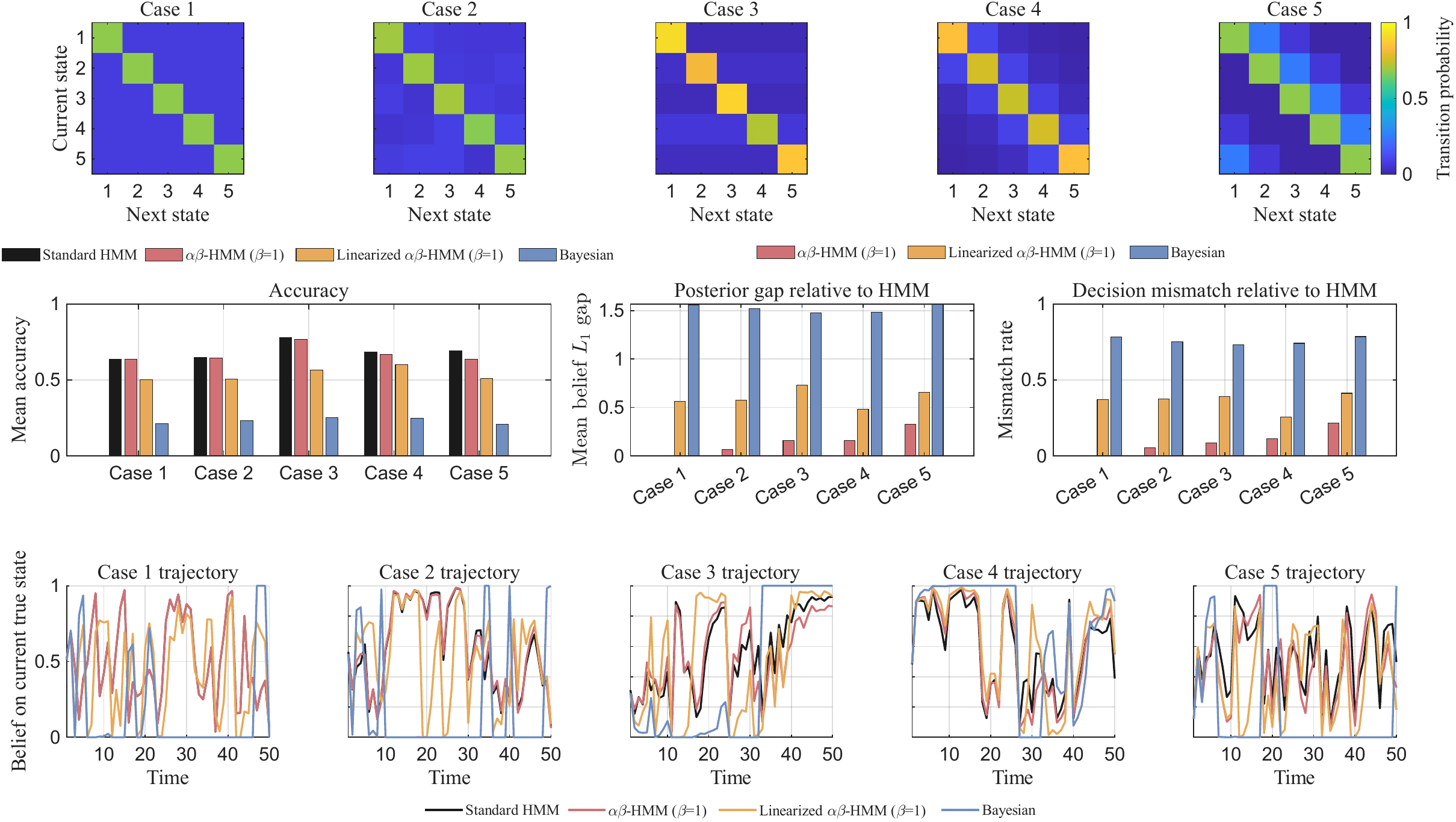}}
\caption{Comparison between the proposed $\alpha\beta$-HMM filter and competing filters under five transition structures. Top row: heatmaps of the five true transition matrices $P^\star$, corresponding to the exact equal-exit, near equal-exit, heterogeneous-exit, local/banded, and directional ring cases, respectively. Middle row: quantitative comparison of the four filters, including the standard HMM with full knowledge of $P^\star$, the proposed $\alpha\beta$-HMM with $\beta=1$, the linearized $\alpha\beta$-HMM with $\beta=1$, and the Bayesian filter. From left to right, the three panels report the mean state-identification accuracy, the mean posterior $L_1$ gap relative to the standard HMM, and the decision mismatch rate relative to the standard HMM. Bottom row: representative trajectories of the belief assigned to the current true state for the five transition cases. The results show that the proposed $\alpha\beta$-HMM remains consistently closer to the standard HMM than the linearized and Bayesian baselines, while the smallest gap is attained in the matched or near-matched equal-exit settings.}
\label{fig:3}
\end{figure*}

In the first numerical experiment, we evaluate how closely the proposed $\alpha\beta$-HMM can approximate the optimal HMM filter under different transition structures. 

We consider a filtering task with $M=5$ candidate states. The observations are generated according to
$\boldsymbol{\xi}_i \sim \mathcal{N}(\theta_i^\star,\sigma_{\mathrm{obs}}^2)$, where the state values are taken as $\theta_m=m+1$, $m=0,\ldots,M-1$, and both the observation-generation noise and the likelihood-model noise are set to $\sigma_{\mathrm{obs}}=\sigma_{\mathrm{like}}=0.8$. The initial prior is chosen to be uniform over the $M$ states. For each experiment, we simulate trajectories of length $T=300$ and average the results over $300$ Monte Carlo runs.

To test the surrogate under different forms of model mismatch, we consider five cases of true transition matrices $P^\star$: 1) an exact equal-exit transition matrix, 2) a near equal-exit transition matrix with small perturbations, 3) a transition matrix with heterogeneous exit rates across states, 4) a local or banded transition matrix that favors nearby states, and 5) a directional ring transition matrix that induces structured cyclic movement. These five cases range from a perfectly matched setting to progressively more structured and anisotropic transition laws, providing a systematic way to assess the approximation quality of the equal-exit surrogate.

For each transition structure, we compare four filters: the standard HMM filter in \eqref{hmm}, with full knowledge of $P^\star$; the proposed $\alpha\beta$-HMM in \eqref{alpha_hmm} with $\beta=1$; the Bayesian filter in \eqref{Bayes}; and the linearized $\alpha\beta$-HMM obtained from Remark~\ref{remark1}, whose corresponding belief update is
\begin{equation}\label{linearized_belief_update}
\boldsymbol{\mu}_{i}(\theta_m)=\frac{\boldsymbol{\mu}_{i-1}^{\,1-\alpha M}(\theta_m)L^\beta(\boldsymbol{\xi}_{i}|\theta_m)}{\sum_{n=0}^{M-1}\boldsymbol{\mu}_{i-1}^{\,1-\alpha M}(\theta_n)L^\beta(\boldsymbol{\xi}_{i}|\theta_n)},
\end{equation}
For both $\alpha\beta$-HMM variants, we set $\beta=1$ and choose $\alpha=\bar h/(M-1)$, where $\bar h$ is the average exit probability of $P^\star$, as introduced in \eqref{optimal_alpha}. In this way, the surrogate model preserves the dominant volatility level of the true environment while discarding detailed transition-direction information.

We evaluate the filters using three criteria: average state-identification accuracy, mean posterior $L_1$ gap relative to the standard HMM, and decision mismatch rate relative to the standard HMM. We also plot a representative trajectory of the belief assigned to the current true state to visualize temporal tracking behavior.


The five transition structures and the corresponding quantitative comparisons are summarized in Fig.~\ref{fig:3}. Several conclusions can be drawn. First, in the equal-exit and near equal-exit cases, the proposed $\alpha\beta$-HMM remains very close to the standard HMM, confirming that the surrogate is highly accurate when the true transition law belongs to, or is close to, the equal-exit family. Second, as the transition structure becomes more heterogeneous or anisotropic, the gap to the optimal HMM becomes more visible but remains moderate overall. Third, the largest degradation occurs in the directional ring case, which is consistent with the analysis in Section~\ref{sec:approximationerror}. Across all five cases, the proposed $\alpha\beta$-HMM stays consistently closer to the standard HMM than the linearized and Bayesian baselines, indicating that the main performance loss is not caused by the surrogate itself, but by the loss of nonlinearity.

\subsection{Real-World Filtering on Human Activity Sequences}

\begin{table*}[t]
\centering
\caption{Comparison of five filters on the real-world activity-sequence dataset.}
\label{tab:realworld_filter_comparison_main}
\resizebox{1.9\columnwidth}{!}{
\begin{tabular}{lcccc}
\toprule
Method & Dynamic params. & Online Acc. & True-state belief  & Switch-window Acc.  \\
|rule
Standard HMM & $M(M-1)$ 
& \textbf{0.8177} 
& \textbf{0.8173} 
& 0.7611 (-0.56\%) \\
HSMM & $M(M-1)+M(D_{\max}-1)$ 
& 0.8136 (-0.50\%) 
& 0.8143 (-0.37\%) 
& \textbf{0.7654} \\
Linearized $\alpha\beta$-HMM & $2$ 
& 0.7133 (-12.76\%) 
& 0.6414 (-21.53\%) 
& 0.4867 (-36.42\%) \\
Bayesian & $0$ 
& 0.3040 (-62.82\%) 
& 0.3026 (-62.98\%) 
& 0.1386 (-81.89\%) \\
|rule
\textbf{Proposed $\alpha\beta$-HMM} & $\mathbf{2}$ 
& \textbf{0.8049} (-1.57\%) 
& \textbf{0.8039} (-1.64\%) 
& \textbf{0.7548} (-1.39\%) \\
\bottomrule
\end{tabular}
}
\end{table*}

To further evaluate the proposed method beyond synthetic Markovian environments, we consider a real-world sequential inference task based on the UCI smartphone-based human activity and postural transition dataset\cite{smartphone-based_recognition_of_human_activities_and_postural_transitions_341}. The dataset contains raw inertial measurements collected from smartphone accelerometers and gyroscopes, together with activity labels over time. Since human activity evolution typically exhibits persistence, duration effects, and subject-dependent switching patterns, the latent dynamics in this task are not expected to follow a homogeneous first-order Markov chain exactly. This makes the dataset suitable for assessing the practical effectiveness of the proposed low-parameter surrogate under model mismatch.

We first partition the raw inertial signals into consecutive fixed-length windows and treat each window as one time instant in the filtering recursion. For each window, compact statistical features are extracted from the six inertial channels and modeled by class-conditional Gaussian likelihoods. We compare five filters: the standard HMM with a fitted fixed transition matrix, the proposed $\alpha\beta$-HMM, the linearized $\alpha\beta$-HMM, the Bayesian filter, and an HSMM baseline with explicit duration modeling. For the standard HMM and HSMM, the transition and duration parameters are estimated from the training subjects. For the proposed method, $\alpha$ is initialized from the average exit rate of the fitted HMM and then tuned jointly with $\beta$ on a validation set. The same validation procedure is applied separately to the linearized baseline. Performance is evaluated using online accuracy, average belief assigned to the current true state, and switch-window accuracy after a true state change. The switch-window accuracy is computed by averaging the accuracy over a short window immediately following each underlying true state transition, and is used to quantify the post-change adaptation ability of each filter.

\begin{figure}[!t]
\centerline{\includegraphics[width=\columnwidth]{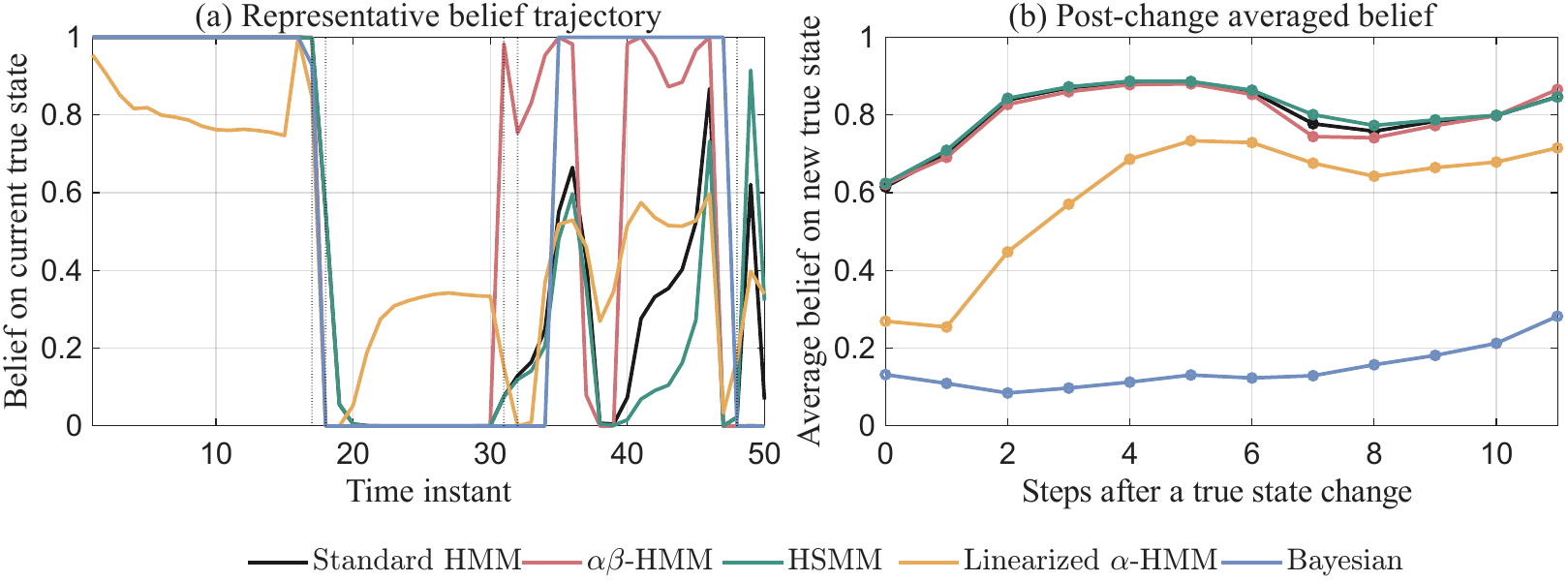}}
\caption{Belief-based comparison of the proposed method and competing filters on the real-world activity sequence dataset. (a) Representative evolution of the belief assigned to the current true state along one test sequence. The vertical dotted lines indicate the true state-change instants. The proposed $\alpha\beta$-HMM tracks the stronger dynamic baselines, namely the standard HMM and HSMM, more closely than the linearized $\alpha$-HMM and the Bayesian filter. (b) Post-change averaged belief on the new true state, obtained by aligning all ground-truth change points and averaging the posterior belief trajectories after each change. The proposed $\alpha\beta$-HMM again remains close to the standard HMM and HSMM, while substantially outperforming the linearized and Bayesian baselines in post-change adaptation.}
\label{fig:5}
\end{figure}

The quantitative comparison in Table~\ref{tab:realworld_filter_comparison_main} shows that the fitted HMM and HSMM achieve the strongest empirical performance, but the proposed $\alpha\beta$-HMM remains very close to both of them on all three reported metrics. In particular, its online accuracy, true-state belief, and switch-window accuracy are all within about $1.6\%$ of the best method in the corresponding column. Here, the column ``Dynamic params.'' counts only the parameters used to model the latent dynamics, since all methods share the same observation model. Under this convention, the standard HMM requires $M(M-1)$ free transition parameters, while the HSMM requires $M(M-1)+M(D_{\max}-1)$ parameters, where $D_{\max}$ denotes the maximum dwell duration, measured in the number of observation windows, used to truncate the explicit duration model. By contrast, the proposed $\alpha\beta$-HMM uses only two interpretable parameters, namely $\alpha$ and $\beta$. Therefore, although it does not explicitly recover the full transition geometry or duration structure, it preserves most of the practical filtering benefit of richer dynamic models with a dramatically simpler parameterization.

This conclusion is further supported by the belief-based visualization in Fig.~\ref{fig:5}. In the representative trajectory, the proposed $\alpha\beta$-HMM tracks the standard HMM and HSMM much more closely than the linearized and Bayesian baselines, and a similar pattern is observed in the post-change averaged belief curve. This indicates that the gain of the proposed method does not come merely from introducing two additional hyperparameters, but from preserving the nonlinear surrogate filtering structure itself. From a broader perspective, the main advantage of the proposed framework is thus not that it strictly outperforms every richer sequence model on a real dataset, but that it offers a competitive filtering rule with low dynamic-model complexity, clear parameter interpretation, and a structure that is much easier to analyze, adapt online, and extend to distributed or multi-agent settings. 

\subsection{Effect of the Exit Probability \texorpdfstring{$\alpha$}{alpha}}

\begin{figure}[!t]
\centerline{\includegraphics[width=\columnwidth]{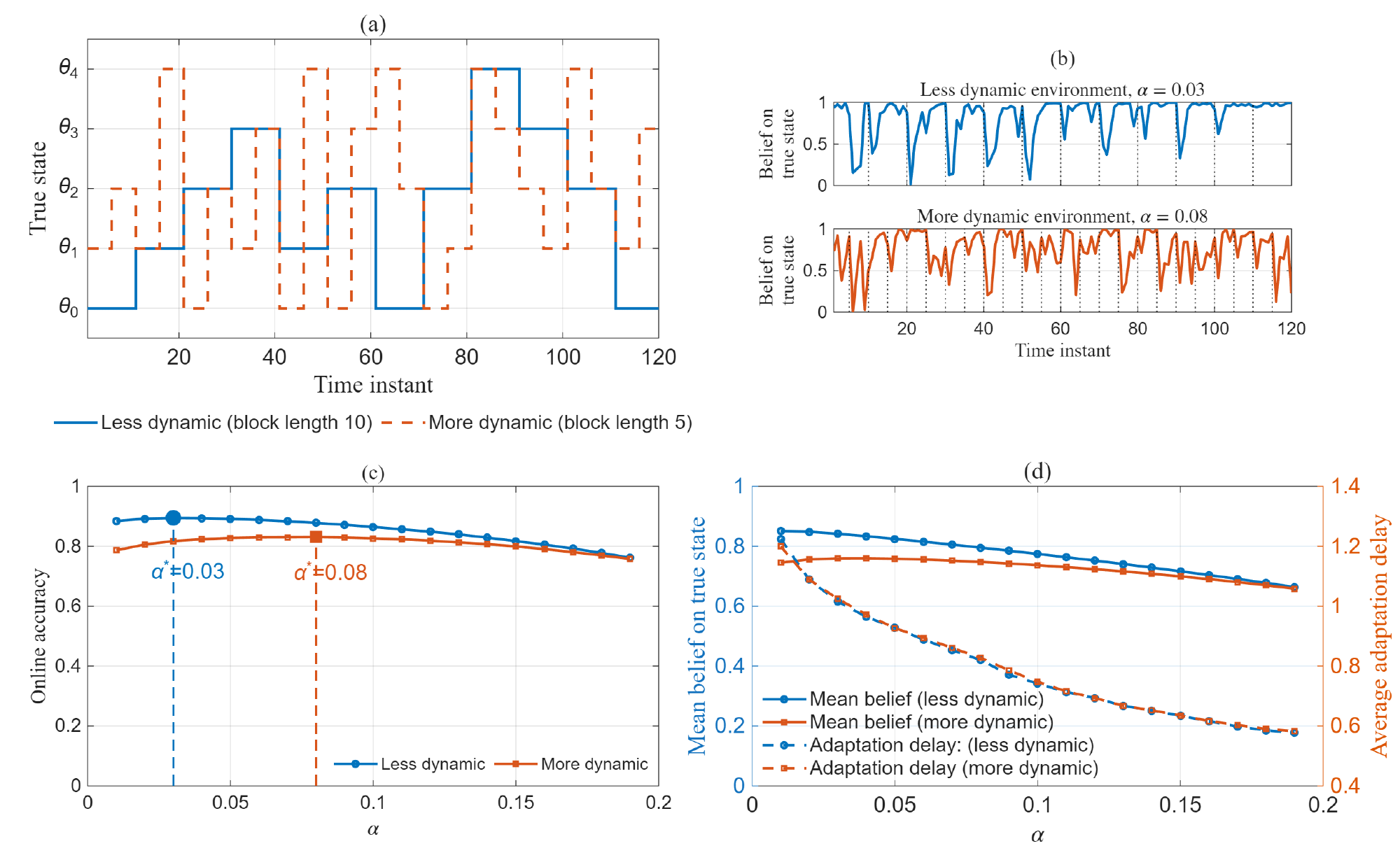}}
\caption{Effect of the exit probability $\alpha$ under two dynamic environments. (a) Representative true-state trajectories in a less and a more dynamic environment, where the underlying true state is reselected every $10$ and $5$ time instants, respectively. (b) Representative belief trajectories of the proposed $\alpha\beta$-HMM in the two environments, obtained using the empirically best values of $\alpha$ for each case. The vertical dotted lines indicate the true state-change instants. (c) Overall online accuracy as a function of $\alpha$ under the two environments. The dashed vertical lines mark the corresponding empirically optimal values of $\alpha$, showing that the more dynamic environment favors a larger $\alpha$. (d) Mean belief on the true state and average adaptation delay as functions of $\alpha$ under the two environments.}
\label{fig:6}
\end{figure}

We next examine how the exit probability $\alpha$ influences the proposed $\alpha\beta$-HMM in dynamic environments. In this experiment, we fix $\beta=1$ in order to isolate the role of $\alpha$. We consider a filtering task with $M=5$ candidate states, where the state values are given by $\theta_m=m+1$, $m=0,\ldots,M-1$. The observations are generated according to $\boldsymbol{\xi}_i \sim \mathcal{N}(\theta_i^\star,\sigma_{\mathrm{obs}}^2)$,
and the likelihood model used by the filter is $L(\boldsymbol{\xi}| \theta_m)=\mathcal{N}(\theta_m,\sigma_{\mathrm{like}}^2)$,
with $\sigma_{\mathrm{obs}}=\sigma_{\mathrm{like}}=0.5$. The initial belief is chosen to be uniform over the $M$ candidate states. We construct two piecewise-constant dynamic environments: a less dynamic one in which the true state is reselected every 10 time instants, and a more dynamic one in which it is reselected every 5 time instants. In both cases, the new state is drawn uniformly from the remaining candidates, so the two environments differ only in their switching frequency. This setting also illustrates that the proposed algorithm is not restricted to strictly Markovian environments.

For each value of $\alpha$, we evaluate overall online accuracy, mean belief on the current true state, and adaptation delay. The latter is defined as the first time instant within the next $H=5$ steps after a state change at which the belief on the new true state exceeds $0.6$; if no such instant exists, the delay is set to $H$. A smaller value of the adaptation delay corresponds to faster post-change adaptation.

The results are shown in Fig.~\ref{fig:6}. It can be seen that the less dynamic environment prefers a smaller value of $\alpha$, while the more dynamic environment favors a larger one. In both cases, increasing $\alpha$ reduces the adaptation delay but also lowers the mean belief on the true state, indicating faster response at the cost of weaker steady-state confidence. As a result, the online accuracy is not monotone in $\alpha$, but peaks at an intermediate value that depends on the dynamic regime. These observations are consistent with Theorems~\ref{thm1} and~\ref{thm2}, and confirm that $\alpha$ provides an explicit and interpretable mechanism for balancing learning quality and adaptation capability.

\subsection{Effect of the Step-Size Parameter \texorpdfstring{$\beta$}{beta}}

\begin{figure}[!t]
\centerline{\includegraphics[width=\columnwidth]{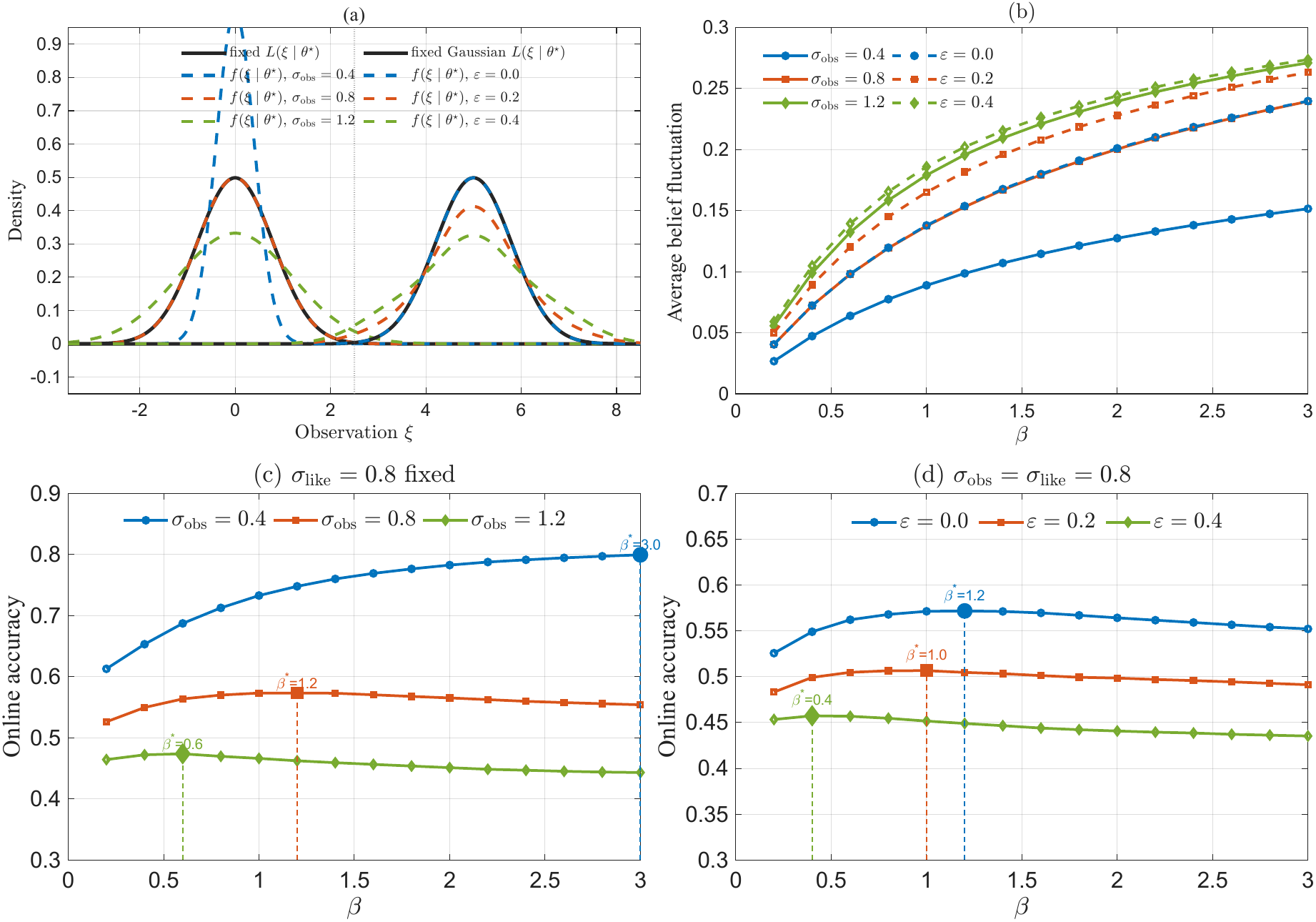}}
\caption{Effect of the step-size parameter $\beta$ under varying observation noise and model mismatch severity. (a) Illustration of the experimental settings. Left: the study of observation noise, where the true observation law $f(\xi|\theta^\star)$ changes with $\sigma_{\mathrm{obs}}$ while the Gaussian likelihood model $L(\xi|\theta^\star)$ is fixed. Right: the study of mismatch severity, where the Gaussian likelihood model remains fixed while the true observation law becomes increasingly contaminated as the contamination probability $\epsilon$ increases. (b) Average belief fluctuation as a function of $\beta$. The solid curves correspond to different observation-noise levels $\sigma_{\mathrm{obs}}$, while the dashed curves correspond to different mismatch severities $\epsilon$. In all cases, a larger $\beta$ leads to stronger fluctuation, and the fluctuation is also larger under higher observation noise or more severe mismatch. (c-d) Overall online accuracy as a function of $\beta$ under two types of conditions. In panel (c), $\sigma_{\mathrm{like}}=0.8$ is fixed while $\sigma_{\mathrm{obs}}$ varies. In panel (d), $\sigma_{\mathrm{obs}}=\sigma_{\mathrm{like}}=0.8$ is fixed while the contamination probability $\epsilon$ varies. The dashed vertical lines mark the empirically optimal values of $\beta$, showing that cleaner observations or a more accurate model favor a larger $\beta$.}
\label{fig:7}
\end{figure}

We next investigate how the step-size parameter $\beta$ influences the proposed $\alpha\beta$-HMM. In this experiment, we fix $\alpha=0.08$ to isolate the role of $\beta$. We consider $M=4$ candidate states with state values $\theta\in\{0,0.7,1.4,2.1\}$. The underlying true state is piecewise constant and is reselected every $6$ time instants from the $M$ candidates, subject to the constraint that it differs from the previous one. 

To examine the role of $\beta$, we consider two sets of experiments. In the first set, we study the effect of observation noise. The observations are generated according to
$\boldsymbol{\xi}_i \sim \mathcal{N}(\theta_i^\star,\sigma_{\mathrm{obs}}^2)$,
while the filter uses the Gaussian likelihood model $L(\boldsymbol{\xi}| \theta_m)=\mathcal{N}(\theta_m,\sigma_{\mathrm{like}}^2)$, with $\sigma_{\mathrm{like}}=0.8$ fixed and $\sigma_{\mathrm{obs}}\in\{0.4,0.8,1.2\}$. In the second set, we study the effect of model mismatch severity. In this case, the true observations are generated from a contaminated model, $f(\boldsymbol{\xi}| \theta_i^\star)
=(1-\epsilon)\mathcal{N}(\theta_i^\star,\sigma_{\mathrm{obs}}^2)
+\frac{\epsilon}{2}\mathcal{N}(\theta_i^\star-\delta_{\mathrm{out}},\sigma_{\mathrm{out}}^2)
+\frac{\epsilon}{2}\mathcal{N}(\theta_i^\star+\delta_{\mathrm{out}},\sigma_{\mathrm{out}}^2)$,
where $\sigma_{\mathrm{obs}}=\sigma_{\mathrm{like}}=0.8$, $\delta_{\mathrm{out}}=1.6$, and $\sigma_{\mathrm{out}}=0.8$. Here, $\epsilon$ denotes the contamination probability, i.e., the probability that an observation is drawn from the outlier component. Therefore, a larger $\epsilon$ corresponds to a more severe model mismatch. Throughout this experiment, however, the filter still assumes the clean Gaussian likelihood model $L(\boldsymbol{\xi}|\theta_m)=\mathcal{N}(\theta_m,0.8^2)$.

For each value of $\beta$, we run the proposed $\alpha\beta$-HMM over multiple Monte Carlo trials and evaluate two performance metrics: the overall online accuracy and the average belief fluctuation. Let $\mathcal{I}_{\mathrm{st}}
=\{i\in\{2,\ldots,T\}:\theta_i^\star=\theta_{i-1}^\star\}$
denote the set of time instants at which the underlying true state remains unchanged. The average belief fluctuation is computed as
\begin{equation}
\frac{1}{|\mathcal{I}_{\mathrm{st}}|}
\sum_{i\in\mathcal{I}_{\mathrm{st}}}
\left|
\mu_i(\theta_i^\star)-\mu_{i-1}(\theta_{i-1}^\star)
\right|.
\end{equation}
This metric measures the sensitivity of the filter to noisy or mismatched observations during stable periods, a larger value indicates that the belief trajectory fluctuates more strongly under the same underlying state.

Figure~\ref{fig:7}(b) shows that the average belief fluctuation increases monotonically with $\beta$, and is further amplified by stronger observation noise or more severe model mismatch. Figures~\ref{fig:7}(c) and~\ref{fig:7}(d) show that the empirically optimal $\beta$ is larger when the observations are cleaner or the model is more accurate. These observations are consistent with Theorems~\ref{thm1}, \ref{thm2}, and \ref{thm3}: larger $\beta$ improves responsiveness and learning when the data are reliable, but also amplifies noise and mismatch, thereby leading to stronger belief fluctuation and degrading performance in less trustworthy settings.

\section{Conclusion and Future Work}

This paper proposed the $\alpha\beta$-HMM as a low-parameter filtering framework for online inference in dynamic environments. The framework addresses the problem from two complementary angles. The parameter $\alpha$ is introduced through an equal exit surrogate to capture environment volatility, while the parameter $\beta$ is introduced through a generalized measurement update to regulate the influence of observational evidence under different levels of observation noise and likelihood model accuracy. In this way, the proposed method provides a compact but interpretable formulation of dynamic filtering that remains closely connected to the structure of HMM-based inference.

A central contribution of this work is that the equal exit surrogate is studied as a modeling principle rather than used merely as a convenient simplification. We analyzed its approximation error, clarified the influence of heterogeneity in the exit probabilities and anisotropy in the destination distributions, and demonstrated through numerical experiments that the resulting low parameter surrogate can remain competitive with the oracle HMM across a broad range of environments. At the same time, the proposed formulation reduces the transition modeling burden from a full $M\times M$ matrix to only two interpretable parameters. This reduction is especially important when the number of candidate states is large, when environment dynamics need to be adapted online, or when one aims to extend the filtering mechanism to network scale and multi agent settings.

Another main contribution is the nonlinear dynamical-systems analysis developed for the proposed recursion. By relating the original stochastic system to a deterministic reference dynamical system, we obtained explicit theoretical results on bounded approximation error, instantaneous error probability, adaptation capability and learning performance. These results also provide principled guidance for parameter selection: $\alpha$ governs the trade-off between memory retention and responsiveness to environmental changes, while $\beta$ controls the influence of observational evidence and should be chosen according to observation quality and likelihood-model accuracy. In particular, larger $\alpha$ improves adaptation at the cost of steady-state confidence, whereas larger $\beta$ is preferable when observations are cleaner and the model is more accurate. Beyond yielding strong guarantees for the proposed filter, this analysis also offers useful insight into the underlying mechanism of HMM-type filtering in changing environments. The numerical experiments further confirmed the qualitative trends predicted by the theory.

Several directions deserve further investigation. On the modeling side, it would be of interest to develop richer low dimensional surrogates that retain more transition structure while preserving analytical tractability. On the algorithmic side, adaptive schemes for tracking time-varying environment volatility and observation reliability may further enhance practical performance. In particular, adaptive updating of the exit probability parameter has already shown promise in our companion study~\cite{11464476}, and extending this perspective to the broader $\alpha\beta$-HMM framework remains an important direction for future work. On the systems side, extending the proposed low parameter filtering framework and its dynamical analysis to distributed and multi-agent networks appears particularly promising. We hope that the present work provides a useful step toward interpretable, low complexity, and theoretically grounded online filtering in dynamic environments.

\bibliographystyle{IEEEtran}
\bibliography{IEEEabrv,main}

\end{document}